\documentclass[conference]{IEEEtran}
\IEEEoverridecommandlockouts
\usepackage{cite}
\usepackage{amsmath,amssymb,amsfonts}
\usepackage{algorithmic}
\usepackage{graphicx}
\usepackage{textcomp}
\usepackage{xcolor}
\usepackage{subcaption}
\usepackage{booktabs}
\usepackage{multirow}
\usepackage{float}
\usepackage[colorlinks=true, linkcolor=black, citecolor=green, urlcolor=blue]{hyperref}
\def\BibTeX{{\rm B\kern-.05em{\sc i\kern-.025em b}\kern-.08em
    T\kern-.1667em\lower.7ex\hbox{E}\kern-.125emX}}
\begin{document}

\title{KAN-SAM: Kolmogorov-Arnold Network Guided Segment Anything Model for RGB-T Salient Object Detection}

\author{
  Xingyuan Li, Ruichao Hou*, Tongwei Ren, Gangshan Wu \\
	State Key Laboratory for Novel Software Technology, Nanjing University\\ Nanjing 210023, China}
\maketitle
\renewcommand{\thefootnote}{}
\footnotetext{*Corresponding author: Ruichao Hou (email: rchou@nju.edu.cn).}

\begin{abstract}
Existing RGB-thermal salient object detection (RGB-T SOD) methods aim to identify visually significant objects by leveraging both RGB and thermal modalities to enable robust performance in complex scenarios, but they often suffer from limited generalization due to the constrained diversity of available datasets and the inefficiencies in constructing multi-modal representations. In this paper, we propose a novel prompt learning-based RGB-T SOD method, named KAN-SAM, which reveals the potential of visual foundational models for RGB-T SOD tasks. Specifically, we extend Segment Anything Model 2 (SAM2) for RGB-T SOD by introducing thermal features as guiding prompts through efficient and accurate Kolmogorov-Arnold Network (KAN) adapters, which effectively enhance RGB representations and improve robustness. Furthermore, we introduce a mutually exclusive random masking strategy to reduce reliance on RGB data and improve generalization. Experimental results on benchmarks demonstrate superior performance over the state-of-the-art methods.
\end{abstract}


\section{Introduction}

\label{sec:intro}
\begin{figure}[t!]
  \centering
  \begin{subfigure}[b]{.5\textwidth} 
      \centering
      \includegraphics[width=\textwidth]{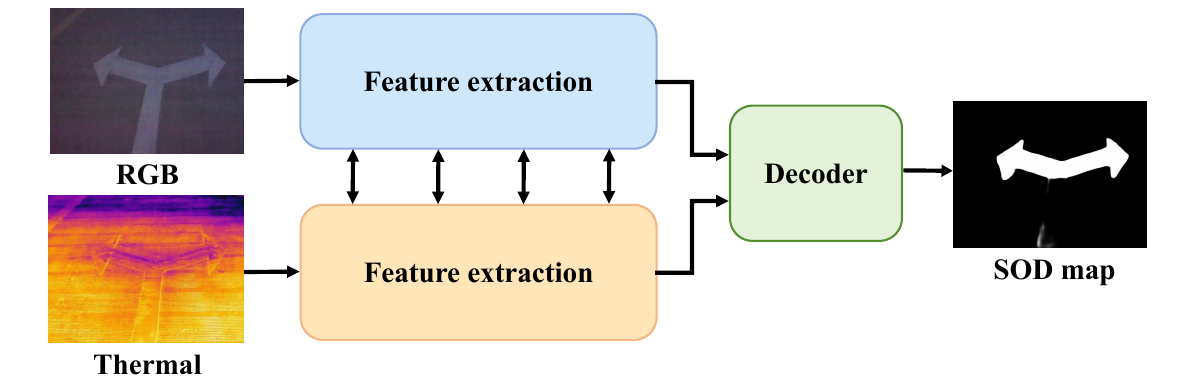} 
      \caption{Existing RGB-T SOD methods}
      \label{fig:sub1}
  \end{subfigure}
  \vspace{0.1cm} 

  \begin{subfigure}[b]{.5\textwidth} 
      \centering
      \includegraphics[width=\textwidth]{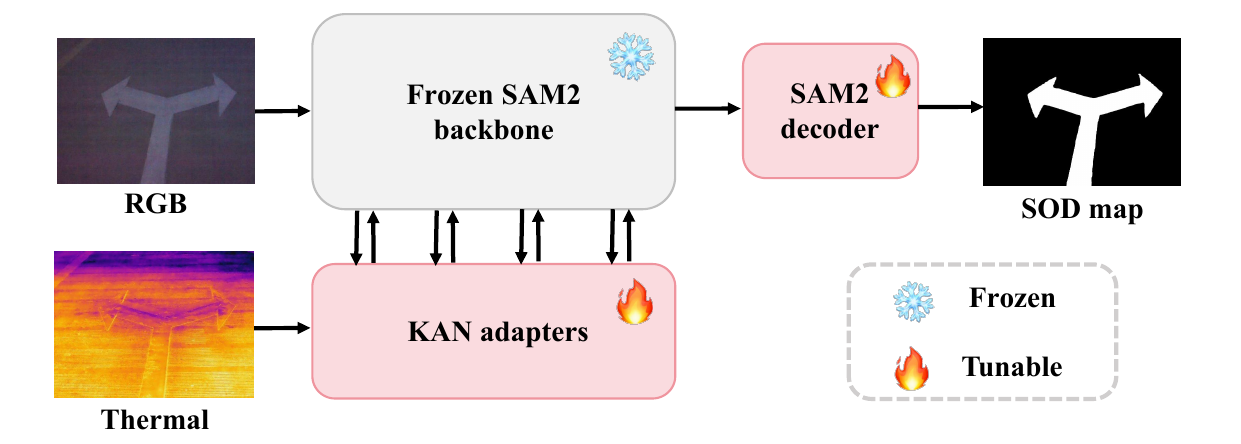} 
      \caption{Our method}
      \label{fig:sub2}
  \end{subfigure}

  \caption{Comparison of frameworks between existing multi-modal methods and our KAN-SAM method.}
  \label{fig:1}
  \vspace{-0.5cm}
\end{figure}

RGB-thermal salient object detection (RGB-T SOD) integrates thermal information with RGB images to improve detection performance under challenging conditions such as low illumination, cluttered environments, and complex backgrounds\cite{ chen2020salient}. 
While traditional RGB-based SOD methods rely solely on visual cues from RGB images, making them less effective in such scenarios\cite{guo2017video}, RGB-T SOD leverages the complementary information provided by thermal data. 
Nevertheless, existing RGB-T SOD methods are limited by the constrained diversity of available datasets and the inefficiencies in constructing multi-modal representations. These methods often suffer from task-specific overfitting and struggle to generalize effectively to unseen environments\cite{huo2021efficient, zhang2019rgb}.
As illustrated in Fig.\ref{fig:1}(a), the SOD map of a road sign generated by existing multi-modal methods fail to accurately capture the complete structure, which arises from inadequate integration of RGB and thermal data, resulting in ineffective feature fusion and decreased performance in complex or low-visibility environments.

Inspired by \cite{gao2024multi}, we propose KAN-SAM, a novel prompt learning-based RGB-T SOD method that extends Segment Anything Model 2 (SAM2)\cite{ravi2024sam} with strong generalization capabilities in diverse downstream tasks\cite{wu2023medical, zhang2024uv}.
Considering SAM2 is designed for RGB-modality data, we introduce prompt learning to enhance the capability of SAM2 by incorporating thermal information as prompts without full parameter fine-tuning.
As shown in Fig.~\ref{fig:1}(b), efficient and accurate Kolmogorov-Arnold Network (KAN) \cite{liu2024kan} adapters are integrated into the SOD framework. KANs outperform MLPs in both accuracy and interpretability: compact KAN models rival or surpass larger MLPs in performance, while their transparent structure enables clear visualization and user-friendly interaction. These adapters effectively infuse thermal prompts into the RGB SOD process, improving robustness and feature abstraction across modalities. Furthermore, we introduce a mutually exclusive random masking strategy during training to reduce over-reliance on RGB data, encouraging the model to better utilize thermal prompts. Finally, the tunable mask decoder of SAM2 is fine-tuned to incorporate task-specific saliency cues, ensuring precise and efficient detection.

The main contributions of our work are as follows:
\begin{itemize}
  \item We propose a novel prompt learning-based RGB-T SOD method that uses KAN adapters to combine RGB and thermal features in SAM2 for RGB-T SOD.
  \item We introduce a mutually exclusive random masking strategy that enhances robust learning of modality-specific features and reduces over-reliance on RGB modality, improving model representations across diverse scenarios.
\end{itemize}

\section{Related Work}

\subsection{RGB-T Salient Object Detection} 

RGB-T SOD leverages both RGB and thermal images to detect salient objects in complex environments. Early methods primarily relied on hand-crafted features like color and texture~\cite{wang2018rgb}. To enhance feature extraction and integration, researchers introduced two-stream networks and cross-modal feature learning methods~\cite{huo2021efficient, zhang2019rgb, wang2021cgfnet}, further improving the robustness of RGB-T SOD models~\cite{xu2024rgb, Zhu_2023_CVPR}. 
In recent years, Transformer-based architectures have made significant progress, such as SwinNet~\cite{liu2021swinnet}, have set new benchmarks for performance. GAN-based methods further enhanced saliency detection by generating high-quality saliency maps~\cite{jiang2020cmsalgan}. 
Unlike these approaches, we designed a novel prompt learning-based RGB-T SOD method built upon foundational models. By leveraging knowledge from large-scale datasets, our method achieves superior performance while reducing the need for extensive training.

\subsection{Segment Anything Model} 

Segment Anything Model (SAM)~\cite{kirillov2023segment} is an innovative framework for image segmentation, known for its exceptional ability to perform prompt-based segmentation on diverse tasks~\cite{chen2023sam}. This versatility has made it widely adopted in fields such as medical image analysis~\cite{ wu2023medical}, motion segmentation~\cite{xie2024moving}, and remote sensing segmentation~\cite{zhang2024uv}. 
Building on the success of SAM, SAM2~\cite{ravi2024sam} brings several important improvements.  It keeps the core prompt-based mechanism but adds enhancements for more complex scenarios. SAM2 is better at handling cluttered environments with overlapping objects. SAM2 is also optimized for real-time processing, making it suitable for time-sensitive applications like live video analysis. Furthermore, it improves segmentation accuracy, especially for high-resolution tasks. Thus, SAM2 expands the scope of image segmentation, making it even more versatile across different fields.

\subsection{Kolmogorov-Arnold Network} 

KANs\cite{liu2024kan} are neural architectures inspired by the Kolmogorov-Arnold theorem, which decomposes multivariate functions into sums of univariate functions. 
Unlike traditional models such as MLPs\cite{rosenblatt1958perceptron}, CNNs\cite{lecun1998gradient}, and RNNs\cite{elman1990finding}, KANs employ learnable and spline-based activation functions instead of fixed ones, enhancing both accuracy and interpretability.  Recent improvements, such as Chebyshev KAN\cite{ss2024chebyshev} and Wav-KAN\cite{bozorgasl2024wav}, have enhanced their ability to approximate nonlinear functions and increased computational efficiency.  Additionally, TKAN\cite{genet2024tkan} has shown better performance in sequential tasks. Given these strengths, KANs are well-suited for design as efficient adapters.

\begin{figure}[t!]
  \centerline{\includegraphics[width=1\textwidth, angle=90]{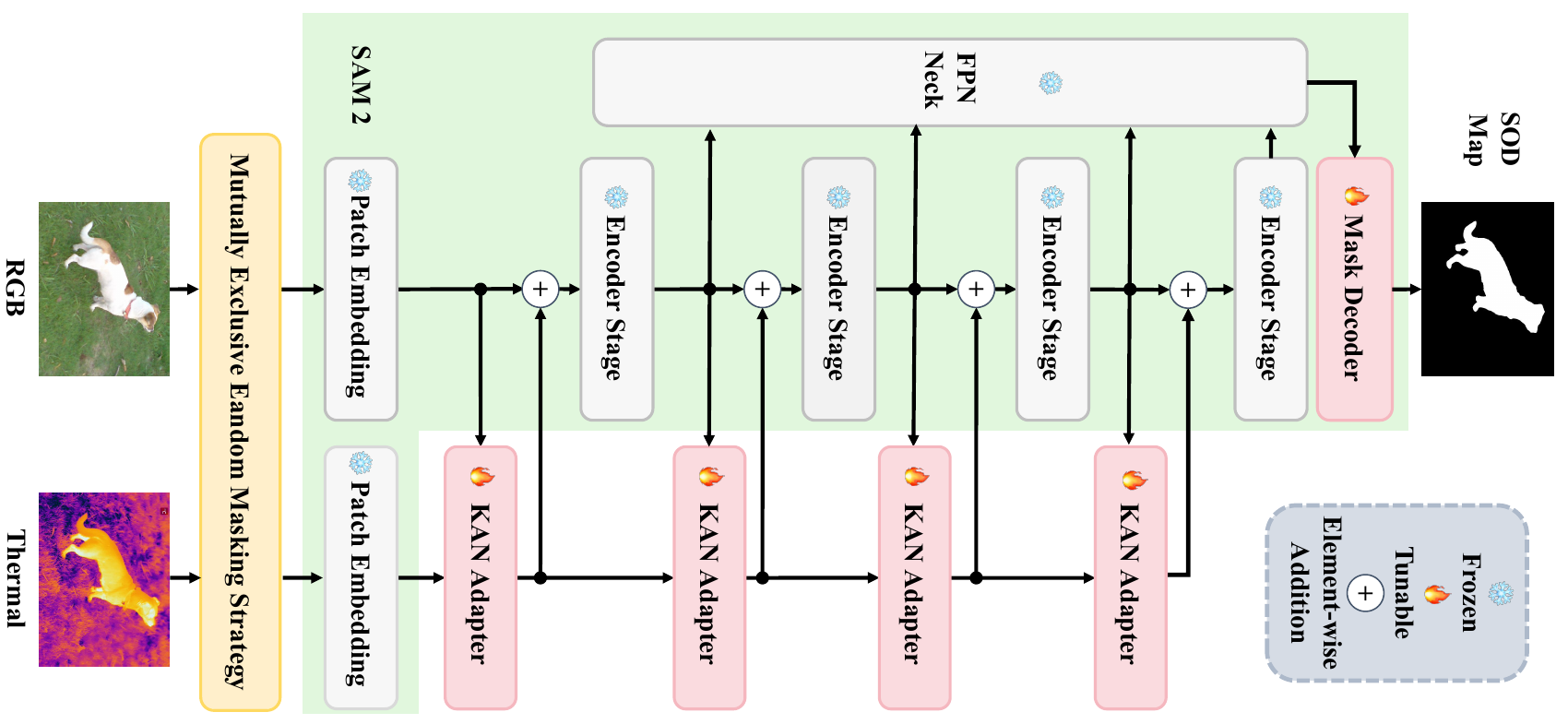}}
  \caption{The framework of our KAN-SAM, which consists of the mutually exclusive random masking strategy, the KAN adapters and the SAM2.}
  \label{fig:model}
  \vspace{-0.3cm}
\end{figure}
\section{Method}
As shown in Fig.~\ref{fig:model}, KAN-SAM builds on the SAM2 framework, keeping most of its components preserved and freezing their parameters. To enable thermal image prompting for RGB images, we integrate KAN adapters. Given a set of RGB and corresponding thermal images, we introduce a mutually exclusive random masking strategy at the input level. The masked images are then converted into embedding features using the patch embedding module. 
The RGB features are processed through the frozen SAM2 Hiera backbone, while the KAN adapters progressively introduce thermal features into the RGB features at each stage. The multi-stage prompting mechanism improves the robustness of multi-modal representations. The multi-scale features extracted in this manner are further refined through the frozen Feature Pyramid Network (FPN) neck for deeper fusion. 
Finally, the fused features are sent to a tunable mask decoder, which generates the final saliency maps. By leveraging multiple KAN adapters for thermal prompting, our method effectively enhances multi-modal feature representations, enabling accurate salient object detection even in complex and challenging scenarios.
\subsection{Mutually Exclusive Random Masking Strategy}
To improve the robustness of KAN-SAM and enable it to learn more discriminative and generalizable features, we introduce a mutually exclusive random masking strategy applied to the input RGB and thermal image pairs. In this strategy, each pixel in the input pair is randomly assigned a 10\% probability of being masked. Importantly, when a pixel is masked, it is masked in only one modality, ensuring mutual exclusivity between the two modalities.

The interleaved masking technique introduces an element of uncertainty and prevents the network from relying entirely on a single modality for information at each pixel location.   By compelling the framework to process incomplete information and compensate for missing data in one modality using the other, the model is trained to utilize the thermal modality as a complementary prompt to enhance the RGB input. This cooperative interaction between the modalities fosters the development of more robust and reliable saliency detection capabilities, particularly under challenging conditions where input quality may be compromised or incomplete.

\subsection{SAM2 Framework}
KAN-SAM builds on SAM2 by incorporating four essential modules: patch embedding, image encoder stages, FPN neck, and mask decoder. The patch embedding, image encoder stages, and FPN neck are frozen to preserve their pre-trained representations, ensuring stability. The mask decoder is tunable, allowing the model to adapt effectively to the saliency detection task.

\textbf{Patch Embedding.}
RGB and thermal images undergo a mutually exclusive random masking strategy before entering the patch embedding module, which projects them into patch embeddings. RGB embeddings are processed through the frozen image encoder stages of SAM2, while thermal embeddings are directed to the KAN adapters, where they are integrated with the RGB embeddings. The KAN adapters act as a prompting mechanism, injecting thermal prompts into the RGB feature flow.

\textbf{Image Encoder.}
The frozen image encoder processes RGB patch embeddings into richer, more semantic features. The KAN adapters inject thermal prompts into this flow by merging thermal and RGB features at each stage, enhancing the feature abstraction process at all hierarchical levels.

\textbf{FPN Neck.}
The FPN Neck fuses multi-scale features from the encoder stages into a unified representation, capturing both coarse and fine details essential for accurate saliency detection. Keeping it unchanged preserves the hierarchical and scale-aware capabilities of the framework.

\textbf{Mask Decoder.}
The tunable mask decoder refines the fused features into final saliency maps. Its adaptability enables the model to capture task-specific information, effectively bridging the gap between the pre-trained SAM2 components and the thermal prompting mechanism, thereby ensuring robust saliency detection.
\begin{figure}[t!]
  \centerline{\includegraphics[width=.5\textwidth]{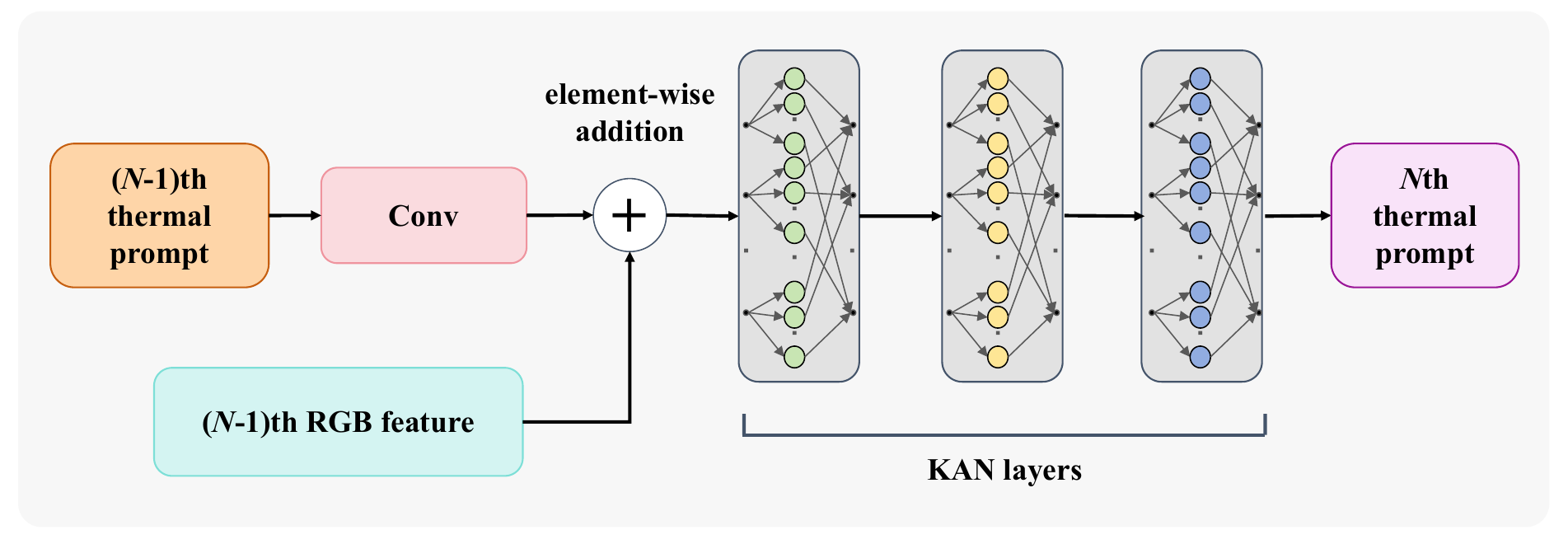}}
  \caption{Detail design of the KAN adapter.}
  \label{fig:kan}
\end{figure}

\subsection{KAN Adapter}

As shown in Fig.~\ref{fig:kan}, we use the KAN adapter, which is an efficient and accurate module designed to dynamically incorporate thermal prompts into the RGB feature flow. It allows the framework to process multi-modal information efficiently, enhancing its overall performance.

The KAN adapter is based on the theoretical foundation of the Kolmogorov-Arnold representation theorem\cite{kolmogorov1957representation}. This theorem states that any continuous multivariate function \( f(x) \) can be expressed as:
\begin{equation}
f(x) = \sum_{q=1}^{2n+1} \Phi_q \left( \sum_{p=1}^n \phi_{q,p}(x_p) \right),
\end{equation}
where \( \phi_{q,p} \) and \( \Phi_q \) are univariate continuous functions, enabling efficient compositional representations.

\textbf{Thermal Prompting.}
The thermal prompting mechanism seamlessly integrates thermal features into the RGB feature flow by injecting thermal cues into the corresponding RGB features both before and after they pass through the KAN adapter. This approach preserves the inherent consistency of the RGB data while refining the overall feature representations. Subsequently, the thermally prompted features are fused with the original RGB features, resulting in a robust, enriched representation that is highly effective for downstream tasks.

To ensure dimensional alignment, the thermal features are transformed using convolutional layers and resized to match the dimensions of the RGB features before processing by the KAN adapter.

\textbf{Feature Transformation.}
The KAN adapter employs multiple layers, each parametrized as a matrix of univariate spline functions:
\begin{equation}
\Phi = \{\phi_{q,p}\}, \quad p = 1, \ldots, n_{\text{in}}, \quad q = 1, \ldots, n_{\text{out}},
\end{equation}
where each \( \phi_{q,p} \) is a learnable spline function. These splines enable the nonlinear transformation and modulation of feature representations.

The forward propagation through a KAN layer is as follows:
\begin{equation}
x_{l+1, j} = \sum_{i=1}^{n_l} \phi_{l, j, i}(x_{l, i}),
\end{equation}
where \( x_{l+1, j} \) represents the activation at the \( j \)-th node of layer \( l+1 \). This value is computed as the sum of the post-activations from the previous layer, enabling efficient and flexible feature transformation.

\textbf{Integration with SAM2 Backbone.}
The KAN adapters are integrated into the SAM2 backbone at multiple stages of the network. These adapters serve as bridges, merging thermal information into the hierarchical feature extraction process. By introducing thermal prompts at each stage, the KAN adapters enable the SAM2 backbone to produce enriched, multi-modal features that improve saliency detection.


\subsection{Loss Function}

Following \cite{li2019dice,jiang2024mirror}, we employ a hybrid loss function that combines Intersection over Union (IoU) loss and Dice loss, balancing region-level and pixel-level accuracy.




The total loss function is defined as follows:
\begin{equation}
\mathcal{L} = \mathcal{L}_{\text{IoU}} + \mathcal{L}_{\text{Dice}}.
\end{equation}

The hybrid loss encourages spatial and structural consistency between the predicted saliency map and the ground truth, ensuring robust and accurate results.

\begin{table*}[h!]

  \setlength\tabcolsep{2pt}

  \scriptsize

    \centering

    \caption{Quantitative comparison of methods using S-measure ($S_m$), MAE, F-measure ($F_{avg}$, $F_{max}$, $F_w$) and E-measure ($E_m$) scores on three RGB-T datasets. The best results and the second best results are highlighted in \textcolor{red}{Red} and \textcolor{blue}{Blue}, respectively.}

      \begin{tabular}{c|cccccc|cccccc|cccccc}

      \hline

      \multirow{2}[1]{*}{Methods} & \multicolumn{6}{c|}{VT5000} & \multicolumn{6}{c|}{VT1000} & \multicolumn{6}{c}{VT821} \\
      \cline{2-19}

      & $F_{avg}\uparrow$ & $F_{max}\uparrow$ & $F_w\uparrow$ & $\emph{MAE}\downarrow$ & $E_m\uparrow$ & $S_m\uparrow$ & $F_{avg}\uparrow$ & $F_{max}\uparrow$ & $F_w\uparrow$ & $\emph{MAE}\downarrow$ & $E_m\uparrow$ & $S_m\uparrow$ & $F_{avg}\uparrow$ & $F_{max}\uparrow$ & $F_w\uparrow$ & $\emph{MAE}\downarrow$ & $E_m\uparrow$ & $S_m\uparrow$ \\

      \hline

      SGDL \cite{tu2019rgb} & 0.672 & 0.737 & 0.558 & 0.089 & 0.824 & 0.750 & 0.764 & 0.807 & 0.652 & 0.090 & 0.856 & 0.787 & 0.731 & 0.780 & 0.583 & 0.085 & 0.846 & 0.764 \\

      CSRNet  \cite{huo2021efficient}& 0.811 & 0.857 & 0.796 & 0.042 & 0.905 & 0.868 & 0.877 & 0.918 & 0.878 & 0.024 & 0.925 & 0.918 & 0.831 & 0.880 & 0.821 & 0.038 & 0.909 & 0.885 \\

      CGFNet  \cite{wang2021cgfnet} & 0.851 & 0.887 & 0.831 & 0.035 & 0.922 & 0.883 & 0.906 & 0.936 & 0.900 & 0.023 & 0.944 & 0.923 & 0.845 & 0.885 & 0.829 & 0.038 & 0.912 & 0.881 \\

      SwinNet \cite{liu2021swinnet} & 0.865 & 0.915 & 0.846 & 0.026 & 0.942 & 0.912 & 0.896 & 0.948 & 0.894 & 0.018 & 0.947 & 0.938 & 0.847 & 0.903 & 0.818 & 0.030 & 0.926 & 0.904 \\

      ADF  \cite{tu2022rgbt}& 0.778 & 0.863 & 0.722 & 0.048 & 0.891 & 0.864 & 0.847 & 0.923 & 0.804 & 0.034 & 0.921 & 0.910 & 0.717 & 0.804 & 0.627 & 0.077 & 0.843 & 0.810 \\

      TNet  \cite{cong2022does}& 0.846 & 0.895 & 0.840 & 0.033 & 0.927 & 0.895 & 0.889 & 0.937 & 0.895 & 0.021 & 0.937 & 0.929 & 0.842 & 0.904 & 0.841 & 0.030 & 0.919 & 0.899 \\

      OSRNet  \cite{huo2022real}& 0.823 & 0.866 & 0.807 & 0.040 & 0.908 & 0.875 & 0.892 & 0.929 & 0.891 & 0.022 & 0.935 & 0.926 & 0.814 & 0.862 & 0.801 & 0.043 & 0.896 & 0.875 \\

      ACMANet \cite{xu2022asymmetric} & 0.858 & 0.890 & 0.823 & 0.033 & 0.932 & 0.887 & 0.904 & 0.933 & 0.889 & 0.021 & 0.945 & 0.927 & 0.837 & 0.873 & 0.807 & 0.035 & 0.914 & 0.883 \\

      MCFNet  \cite{ma2023modal}& 0.848 & 0.886 & 0.836 & 0.033 & 0.924 & 0.887 & 0.902 & 0.939 & 0.906 & 0.019 & 0.944 & 0.932 & 0.844 & 0.889 & 0.835 & 0.029 & 0.918 & 0.891 \\

      CMDBIF \cite{xie2023cross} & 0.868 & 0.892 & 0.846 & 0.032 & 0.933 & 0.886 & 0.914 & 0.931 & 0.909 & 0.019 & \textcolor{blue}{0.952} & 0.927 & 0.856 & 0.887 & 0.837 & 0.032 & 0.923 & 0.882 \\
      MITF-Net\cite{chen2022modality}& 0.880 & 0.899 & 0.870 & 0.025 & 0.943 & 0.910 & 0.915 & 0.938 & 0.906 & 0.016 & 0.949 & 0.938 & 0.865 & 0.891 & 0.853 & 0.027 & 0.927 & 0.905\\

      CAVER \cite{pang2023caver}& 0.856 & 0.897 & 0.849 & 0.028 & 0.935 & 0.899 & 0.906 & 0.945 & 0.912 & 0.016 & 0.949 & 0.938 & 0.854 & 0.897 & 0.846 & 0.026 & 0.928 & 0.897 \\
      ADNet  \cite{fang2023adnet}& \textcolor{blue}{0.893}  & \textcolor{blue}{0.924}  & \textcolor{blue}{0.884}  & \textcolor{blue}{0.022}  & \textcolor{blue}{0.953}  & \textcolor{blue}{0.924}  & 0.916  &\textcolor{red}{0.952}  & 0.920  &\textcolor{blue}{0.015}  & \textcolor{blue}{0.952}  & \textcolor{blue}{0.944}  & 0.869  & \textcolor{red}{0.915}  & 0.860  & \textcolor{red}{0.024}  & 0.930  & \textcolor{red}{0.915}  \\
      WGOFNet  \cite{wang2024weighted}& 0.883  & 0.912  & 0.873  & 0.025  & 0.945  & 0.911  & \textcolor{blue}{0.919}  & {0.946}  & \textcolor{blue}{0.922}  & 0.016  & 0.951  & 0.940  & \textcolor{blue}{0.875}  & \textcolor{blue}{0.911}  & \textcolor{blue}{0.868}  & \textcolor{blue}{0.025}  & \textcolor{red}{0.934}  & \textcolor{blue}{0.908}  \\
      UMINet  \cite{gao2024uminet} & 0.831  & 0.877  & 0.820  & 0.035  & 0.919  & 0.882  & 0.892  & 0.935  & 0.896  & 0.021  & 0.941  & 0.926  & 0.791  & 0.849  & 0.782  & 0.054  & 0.879  & 0.858  \\
      \hline
      \textbf{Ours}      &\textcolor{red}{0.909} & \textcolor{red}{0.931} & \textcolor{red}{0.905} & \textcolor{red}{0.020} & \textcolor{red}{0.957} & \textcolor{red}{0.927} &
      \textcolor{red}{0.930} & \textcolor{blue}{0.947} & \textcolor{red}{0.934} & \textcolor{red}{0.013} & \textcolor{red}{0.958} & \textcolor{red}{0.946} &
      \textcolor{red}{0.883} & \textcolor{blue}{0.911} & \textcolor{red}{0.880} &\textcolor{blue}{0.025} & \textcolor{blue}{0.932} & \textcolor{red}{0.915}\\

      \hline
      \end{tabular} 
    \label{tab:1}
  \end{table*}
\section{Experiments}
  \begin{figure*}[t!]
    \centerline{\includegraphics[width=.98\textwidth]{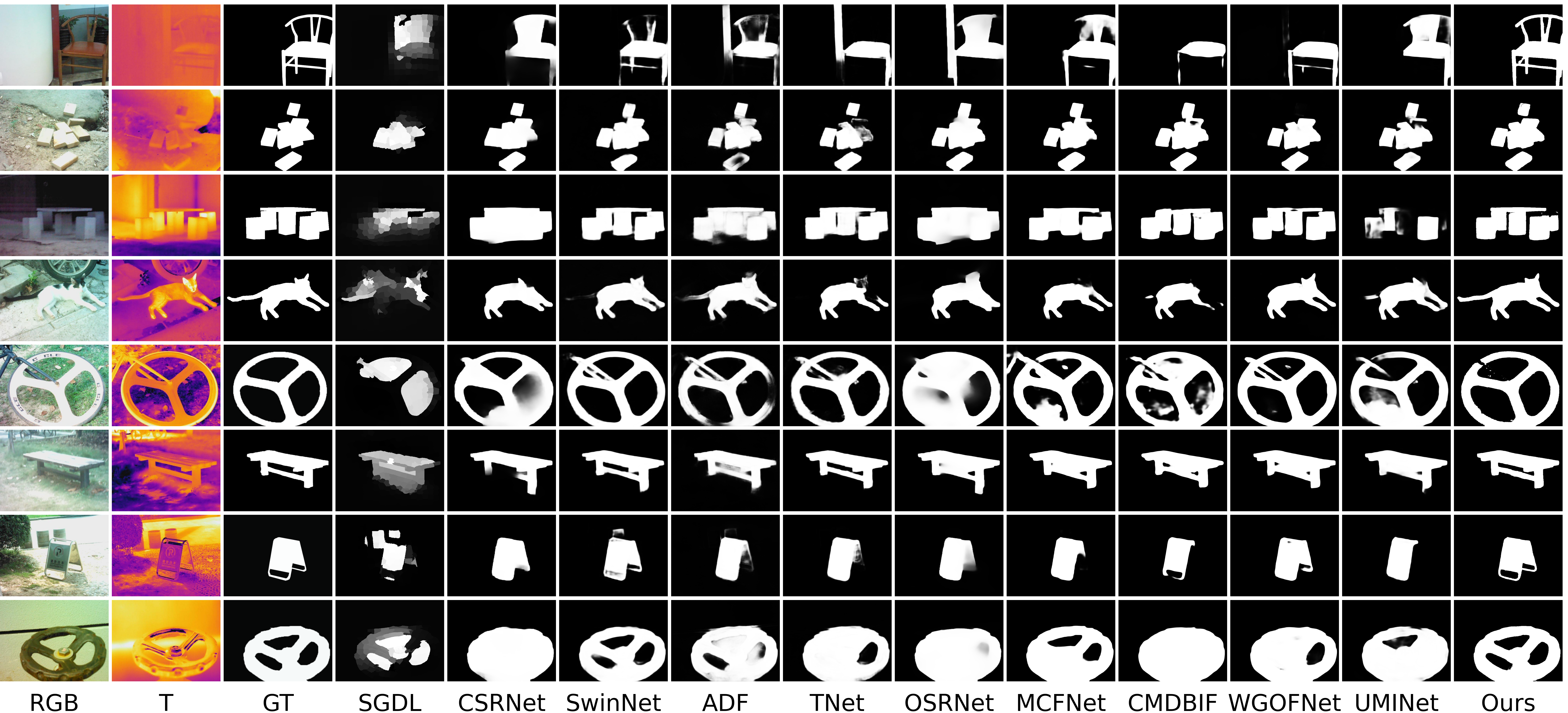}}
    \caption{Visualization comparison of KAN-SAM with the state-of-the-art methods.}
    \label{fig:map}
  \end{figure*}

  \begin{table*}[t!]

    \setlength\tabcolsep{2pt}
  
    \scriptsize
  
      \centering
  
      \caption{Ablation studies on RGB-T SOD datasets. The best results are highlighted in \textbf{Bold}.}
  
        \begin{tabular}{c|cccccc|cccccc|cccccc}
  
        \hline
  
        \multirow{2}[1]{*}{Methods} & \multicolumn{6}{c|}{VT5000} & \multicolumn{6}{c|}{VT1000} & \multicolumn{6}{c}{VT821} \\
        \cline{2-19}
        & $F_{avg}\uparrow$ & $F_{max}\uparrow$ & $F_w\uparrow$ & $\emph{MAE}\downarrow$ & $E_m\uparrow$ & $S_m\uparrow$ & $F_{avg}\uparrow$ & $F_{max}\uparrow$ & $F_w\uparrow$ & $\emph{MAE}\downarrow$ & $E_m\uparrow$ & $S_m\uparrow$ & $F_{avg}\uparrow$ & $F_{max}\uparrow$ & $F_w\uparrow$ & $\emph{MAE}\downarrow$ & $E_m\uparrow$ & $S_m\uparrow$ \\
  
        \hline
        SAM2&0.886 & 0.911 & 0.877 & 0.026 & 0.945 & 0.910 & 0.910 & 0.930 & 0.908 & 0.020 & 0.945 & 0.928 & 0.864 & 0.895 & 0.860 & 0.031 & 0.923 & 0.903\\
  
        SAM2+Mask&0.878 & 0.910 & 0.872 & 0.026 & 0.942 & 0.909 & 0.900 & 0.930 & 0.907 & 0.021 & 0.942 & 0.930 & 0.849 & 0.890 & 0.849 & 0.033 & 0.913 & 0.899     \\
  
        SAM2+KAN &0.899 & 0.926 & 0.896 & {0.021} & 0.953 & 0.924 & 0.918 & 0.941 & 0.923 & {0.015} & 0.951 & 0.940 & 0.873 & 0.909 & 0.872 & 0.027 & 0.928 & 0.911        \\
        SAM2+Mask+KAN  &\textbf{0.909} & \textbf{0.931} & \textbf{0.905} & \textbf{0.020} & \textbf{0.957} & \textbf{0.927} &
        \textbf{0.930} & \textbf{0.947} & \textbf{0.934} & \textbf{0.013} & \textbf{0.958} & \textbf{0.946} &
        \textbf{0.883} & \textbf{0.911} & \textbf{0.880} & \textbf{0.025} & \textbf{0.932} & \textbf{0.915}\\\hline
        \end{tabular} 
      \label{tab:2}
    \end{table*}
    \begin{table}[t!]
      \centering
      \caption{Parameter counts and complexity comparison.}
      \label{tab:table}
          \resizebox{0.48\textwidth}{!}{
          \large
          \begin{tabular}{*{10}{c}}
              \toprule
              Methods	&Params (M)	& Fine-tune Params (M) &FLOPs (T)\\
              \midrule
              SAM2 (Original)	&641.720	&641.720&1.821\\
              SAM2 with MLP &644.502	&2.782 (MLP) + 4.213 (SAM2 decoder) &1.827\\
              \textbf{SAM2 with KAN (Ours)}&\textbf{643.588}	& \textbf{1.868 (KAN) + 4.213 (SAM2 decoder)} &\textbf{1.824}\\
              \bottomrule
          \end{tabular}
          }
          \vspace{-0.5cm}
    \end{table}
  \subsection{Datasets and Experimental Settings}
  
  We use 2,500 image pairs from the VT5000 dataset~\cite{tu2022rgbt} for training. The remaining 2,500 image pairs from VT5000, along with the full VT821~\cite{wang2018rgb} and VT1000~\cite{tu2019rgb} datasets, are used for testing. The VT821 dataset contains 821 image pairs, while VT1000 includes 1,000 image pairs. All these image pairs consist of highly aligned RGB and thermal images.
  For evaluation, we adopt the metrics used in MITF-Net~\cite{chen2022modality}, \emph{i.e.}, $F_{avg}$, $F_{max}$, $F_w$, \emph{MAE}, $E_m$, and $S_m$. These metrics provide a comprehensive assessment of the performance of different methods.
  
  KAN-SAM is implemented using the PyTorch framework, with all experiments conducted on two NVIDIA RTX 4090 GPUs.
  To support SAM2, the input images are resized to a resolution of 512 × 512. Data augmentation is introduced to the training set using random flipping, random rotation, and random cropping. The framework is trained using the AdamW optimizer with an initial learning rate of 1e-4. The batchsize is set to 8, the weight decay is set to 5e-4, the momentum is set to 0.9, the gradient clipping value is set to 0.5 and the maximum number of training epochs is fixed at 30. The complete training process takes approximately 6 hours.

  \subsection{Comparison with the State-of-the-Art}

  We compare KAN-SAM with the state-of-the-art methods, namely SGDL~\cite{tu2019rgb}, CSRNet~\cite{huo2021efficient}, CGFNet~\cite{wang2021cgfnet}, SwinNet~\cite{liu2021swinnet}, ADF~\cite{tu2022rgbt}, TNet~\cite{cong2022does}, OSRNet~\cite{huo2022real}, ACMANet~\cite{xu2022asymmetric}, MCFNet~\cite{ma2023modal}, CMDBIF~\cite{xie2023cross}, MITF-Net~\cite{chen2022modality}, CAVER~\cite{pang2023caver}, ADNet~\cite{fang2023adnet}, WGOFNet~\cite{wang2024weighted}, and UMINet~\cite{gao2024uminet}. 
  We evaluate the performance of KAN-SAM and these methods on three typical datasets for RGB-T SOD task, \emph{i.e.}, VT5000, VT1000, and VT821. 
  
  \textbf{Quantitative Analysis.}
  Table~\ref{tab:1} shows that our proposed KAN-SAM consistently outperforms existing methods across multiple evaluation metrics on the VT5000, VT1000, and VT821 datasets. Specifically, on the VT5000 dataset, KAN-SAM achieves the highest scores in $F_{avg}$, $F_{max}$, $F_w$, MAE, $E_m$, and $S_m$, surpassing the second-best method, ADNet, by up to 1.6\% in $F_{avg}$ and reducing MAE by 0.002. Similarly, on the VT1000 dataset, KAN-SAM leads in $F_{avg}$, $F_w$, MAE, $E_m$, and $S_m$, outperforming the second-best method by 1.2\% as much as in $F_w$ and decreasing MAE by 0.002. For the VT821 dataset, KAN-SAM secures the top position in $F_{avg}$, $F_w$, and $S_m$, with improvements of up to 1.2\% in $F_w$ compared to the second-best methods. These consistent advancements are attributed to our innovative network architecture and the effective integration of RGB and thermal modalities, which enable the model to leverage complementary information for accurate and robust salient object detection.
  
  \textbf{Qualitative Analysis.}
  As shown in Fig.~\ref{fig:map}, KAN-SAM demonstrates superior visual performance compared to state-of-the-art methods. In the second row, it accurately detects the salient object with sharp edges and clear contours, in contrast to TNet, which yields blurred boundaries. In the first and third rows, KAN-SAM maintains object structures and intricate details under challenging conditions like complex shapes or poor lighting, whereas CSRNet struggles in low-light environments, resulting in incomplete and noisy saliency maps. The fifth row demonstrates that KAN-SAM effectively removes irrelevant backgrounds and enhances contrast in cluttered scenes, leading to high-quality saliency maps. In contrast, CMDBIF fails to separate objects from backgrounds well, causing more noise and lower clarity. These examples show that KAN-SAM consistently produces saliency maps that closely match the ground truth across various complex situations.
  
  
  \subsection{Ablation Study}
  
  The ablation study evaluates the contributions of mutually exclusive random masking strategy and the KAN adapter. Without any additional modifications, SAM2 already establishes a strong baseline. However, incorporating the mutually exclusive random masking strategy alone results in a slight performance decline, as the framework loses information from masked regions without complementary guidance, which hinders generalization. 
  In contrast, integrating the KAN adapter significantly improves performance by incorporating thermal information, enhancing the ability of the model to detect salient objects. Moreover, the combination of the mutually exclusive random masking strategy with the KAN adapter yields the best results: the masking strategy prevents over-reliance on RGB data, while the KAN adapter leverages thermal prompts to compensate for the missing information. 
   
  \subsection{Complexity Analysis}
  Our work leverages SAM2's superior performance while addressing efficiency through prompt learning.
  As shown in Table~\ref{tab:table}, our method only updates 0.95\% of parameters (6.081M/641.720M) compared to full fine-tuning.
  Moreover, our framework can be further optimized for efficiency, as the SAM2 backbone can be replaced with lightweight alternatives. Exploring this direction will be the focus of our future work.
  Table~\ref{tab:table} also demonstrates that KAN adapters require only 1.868M parameters (32.8\% reduction from MLP adapters' 2.782M) and 3GFLOPs (50\% reduction from MLP adapters' 6GFLOPs).
  This stems from KAN's spline-based architecture, which reduces redundancy while maintaining nonlinear approximation capabilities.

\section{Conclusion}
In this paper, we proposed a novel prompt learning-based RGB-T SOD method KAN-SAM, which leverages efficient and accurate KAN adapters to incorporate thermal features as guiding prompts. The thermal prompted KAN adapters improved multi-modal representations, enhancing robustness in challenging scenarios like low illumination and cluttered scenes. Additionally, we employed a mutually exclusive random masking strategy to reduce dependence on RGB data. Experimental results on various benchmarks show that KAN-SAM outperforms state-of-the-art methods, underscoring the potential of visual foundational models for advancing RGB-T saliency object detection.

\section*{Acknowledgment}
This work was supported by the National Natural Science Foundation of China (62072232), the Key R\&D Project of Jiangsu Province (BE2022138), the Fundamental Research Funds for the Central Universities (021714380026), and
the Collaborative Innovation Center of Novel Software Technology and Industrialization.
\bibliographystyle{IEEEbib}
\bibliography{icme2025references}

\end{document}